\documentclass[letterpaper,pra,twocolumn,showpacs,superscriptaddress]{revtex4} 
\usepackage[mathscr]{eucal}
\usepackage{amsmath}
\usepackage{amssymb}
\usepackage{amsfonts}
\usepackage{ae}
\usepackage{epsfig}
\usepackage{subfigure}
\usepackage{caption}

\usepackage{graphicx}
\usepackage{dcolumn}
\usepackage{bm}
\DeclareGraphicsExtensions{.eps}

\begin{document}
\title{Multimode quantum properties of a self-imaging OPO: \\squeezed vacuum and EPR beams generation}

\author{L. Lopez} \affiliation{Laboratoire Kastler Brossel, Universit\'e Pierre et Marie-Curie-Paris 6, ENS, CNRS ; 4 place Jussieu, 75005 Paris, France}
\author{B. Chalopin}\affiliation{Laboratoire Kastler Brossel, Universit\'e Pierre et Marie-Curie-Paris 6, ENS, CNRS ; 4 place Jussieu, 75005 Paris, France}
\author{A. Rivi\`ere de la Souch\` ere} \affiliation{Laboratoire Kastler Brossel, Universit\'e Pierre et Marie-Curie-Paris 6, ENS, CNRS ; 4 place Jussieu, 75005 Paris, France}
\author{C. Fabre} \affiliation{Laboratoire Kastler Brossel, Universit\'e Pierre et Marie-Curie-Paris 6, ENS, CNRS ; 4 place Jussieu, 75005 Paris, France}
\author{A. Ma\^\i tre} \affiliation{Laboratoire Kastler Brossel, Universit\'e Pierre et Marie-Curie-Paris 6, ENS, CNRS ; 4 place Jussieu, 75005 Paris, France}\affiliation{Institut des NanoSciences de Paris, Universit\'e Pierre et Marie Curie-Paris 6, Campus Boucicaut, 140 rue de Lourmel, 75015 Paris, France}
\author{N. Treps}\email{treps@spectro.jussieu.fr}\affiliation{Laboratoire Kastler Brossel, Universit\'e Pierre et Marie-Curie-Paris 6, ENS, CNRS ; 4 place Jussieu, 75005 Paris, France}

\date{\today}
\begin{abstract}
  We investigate the spatial quantum properties of the light emitted
  by a perfectly spatially degenerate optical parametric oscillator
  (self-imaging OPO). We show that this device produces local
  squeezing for areas bigger than a coherence are that depends on the
  crystal length and pump width. Furthermore, it generates local EPR
  beams in the far field. We show, calculating the eigenmodes of the
  system, that it is highly multimode for realistic experimental
  parameters.
\end{abstract}
\pacs{42.50.Dv, 42.65.Yj, 42.60.Da}

\maketitle

\section{Introduction}


Highly multiplexed quantum channels are more and more needed as
complexity increases in the quantum communication and information
protocols.  They can be obtained by coupling many single mode quantum
channels \cite{Furosawa}, but also by directly using highly multimode
quantum systems. In addition, the resolution of several problems in
quantum imaging \cite{book quantum imaging} requires the generation of
non-classical states of light having adjustable shapes in the
transverse plane: this is the case for superresolution
\cite{KolobovResolution}, or for image processing below the standard
quantum noise level \cite{trepsdelaubert}. For all these reasons, it
is very important to develop a source of highly multimode non
classical light (squeezed and/or entangled) of arbitrary transverse
shape.

In the continuous variable regime, where optical resonators are
necessary to efficiently produce non-classical states, one of the keys
to successfully generate multimode light is the ability to operate a
multimode optical resonator. Indeed, many theoretical proposals rely
on the use of an Optical Parametric Oscillator (OPO) operated below
threshold with planar cavities \cite{kolobovlugiato95} or with confocal cavities
\cite{Grangier} \cite{Mancini} \cite{Petsas} which spatially filter half of the
transverse modes. However, these proposals still did predict the
arising of local vacuum squeezing and image amplification. Hence, we
propose here to keep the parametric process to generate
non-classical light but also to overcome the problems encountered by
the use of a full transverse degenerate cavity : the self-imaging
cavity \cite{Arnaud}. This type of resonator, used for instance to
improve the power of multimode lasers \cite{Couderc}, is in principle able
to transmit any optical image within its spatial bandwidth.

The aim of this article is to demonstrate that the self-imaging OPO is an excellent candidate to produce local squeezing, image amplification
and also local EPR beams, taking into account its physical limitations such as the thickness of the crystal and the finite size of the various
optical beams and detectors.

The following section (section \ref{sec2}) describes the experimental
configuration and develops the theoretical model, as well as the
method used to determine the squeezing spectra measured in
well-defined homodyne detection schemes. In section \ref{sec3}, the
results for such quantities respectively in the near field and in the
far field are given, and we investigate the generation of \textit{EPR}
beams. Finally, in section \ref{sec4} we compute the eigenmodes of the
system and show that they are closed to Hermite Gauss modes.

\section{Self-imaging Optical Parametric Oscillator}
\label{sec2}
\subsection{The self-imaging cavity}
\begin{figure}[h!]
\centerline{ \scalebox{.4}{\includegraphics*{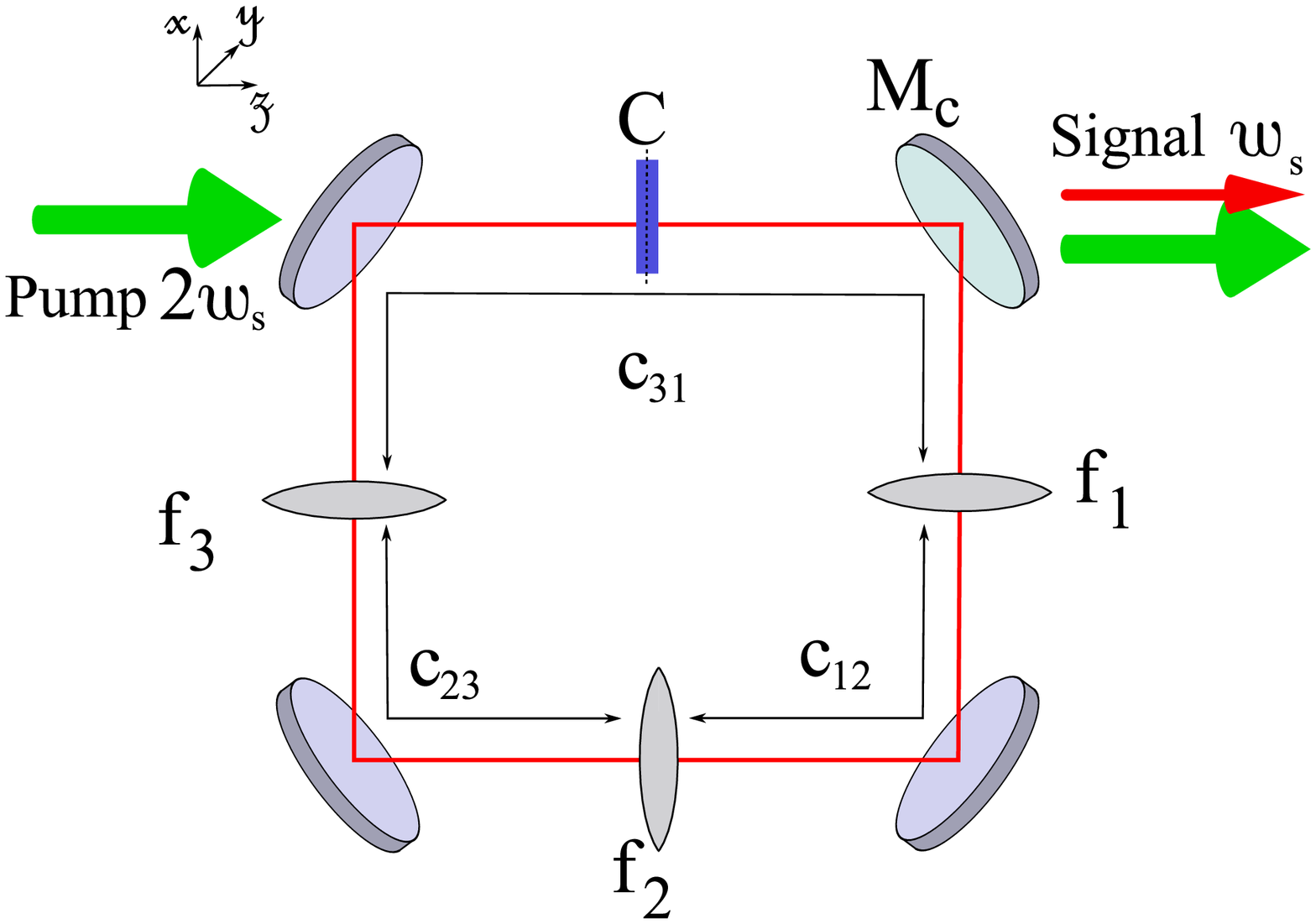}} }
\caption{ Self-imaging OPO scheme in a ring cavity
  configuration} \label{fig:fig2}
\end{figure}

We consider the parametric down conversion taking place in a
self-imaging optical parametric oscillator whose cavity has been
depicted in the pioneer article of $Arnaud$ \cite{Arnaud}. Such a
cavity is a fully transverse degenerate one, which implies that all
the transverse modes of same frequency resonate for the same cavity
length. From a geometrical point of view an optical cavity is self
imaging when an arbitrary ray retraces its own path after a single
round trip. In such a cavity, in the paraxial approximation, the
$ABCD$ matrix $M$ after one round trip is equal to identity:
\begin{equation}\label{ABCD}
\begin{pmatrix}
  A & B \\
  C & D
\end{pmatrix}=
\begin{pmatrix}
  1 & 0 \\
  0 & 1
\end{pmatrix}
\end{equation}
The simplest self imaging ring cavity requires three lenses of focal
length $f_{i}$, $i=1,2,3$ (\cite{Arnaud}). As depicted in
Fig(\ref{fig:fig2}), the ring cavity is self-imaging provided the
distances $c_{i j}$ of the image plane of the lens $i$ and the objet
plane of the lens $j$ are given by:
\begin{equation}\label{condition}
c_{12}=\frac{f_{1}f_{2}}{f_{3}}, \\
c_{23}=\frac{f_{2}f_{3}}{f_{1}}, \\
c_{31}=\frac{f_{3}f_{1}}{f_{2}}
\end{equation}

Let us consider an Optical Parametric Oscillator (OPO) whose cavity is
the self-imaging one described in Fig.\ref{fig:fig2} (\cite{Mancini}).
A type I parametric medium of length $l_c$ is centered on the plane C
located at the longitudinal coordinate z=0. The OPO is pumped by a
gaussian $TEM_{00}$ field $E_p$ of amplitude $A_p$ and frequency
$\omega_p=2\omega_s$. Its waist $w_p$ is located at the plane C. The
OPO works in a longitudinal degenerate operation for which signal and
idler have the same frequency $\omega_s$. We assume that for the pump
wave, all the mirrors are totally transparent, and that for the signal
field, the coupling mirror $M_c$ has a small transmission $t$, the
other three mirrors being perfectly reflecting.

\subsection{Electric field operators}

We will follow an operational approach \cite{GattiLugiato} close to the one developed in the confocal
case \cite{Lopez}, and in order to keep the present article concise we
give only the main steps of the calculation. The intracavity signal
field at frequency $\omega_s$ is described by a field envelope
operator ${B}(\textbf{x},z)$. In the self imaging resonator, at
resonance, the field can be decomposed on any transverse mode basis
(such as the Gauss-Laguerre modes for instance). The field operator
becomes :
\begin{equation}
{B}(\textbf{x},z,t)=\sum_{l}
f_{l}(\textbf{x},z){a}_{l}(z,t)\, ,
\label{expansion}
\end{equation}
where ${a}_{l}(z,t)$ is the annihilation operator of a photon in mode
$l$ at the cavity position $z$ and at time $t$. $f_{l}$ is the
amplitude of the $l$ mode. This field obeys the standard equal time
commutation relation at a given transverse plane at position $z$:
\begin{equation} [{B}(\textbf{x},z,t),{B}^\dagger
(\textbf{x'},z,t)]=\delta(\textbf{x}-\textbf{x'}) .
\end{equation}
Indeed, contrary to the confocal case \cite{Lopez} or to any partially
imaging cavity case, this operator is the same as the one in the
vacuum as no spatial filtering is induced by the cavity. In the regime
below threshold considered here, the pump is not depleted, and
fluctuations of the pump field do not contribute, at first order to
the fluctuations of the signal.

The interaction Hamiltonian of the system, taking into account the
thickness of the crystal and the shape of the pump, is given by
\begin{eqnarray}
H_{int} & = & \frac{i\hbar g}{2 l_c}\int_{-l_c/2}^{l_c/2}dz'\int\int d^2
x' \{A_P(\textbf{x'},z')[{B^{\dagger}}(\textbf{x'},z',t)]^2 \nonumber \\
& & - h.c.\} \label{Hamiltonian} \, ,
\end{eqnarray}
where g is the coupling constant proportional to the second order
nonlinear susceptibility $\chi^{(2)}$.

\subsection{Evolution equation of the field }
In this section, we investigate the intracavity evolutions of the signal field in the crystal plane C (near field) and of its spatial fourier
transform (far field), taking into account the crystal thickness and the finite size of the pump. The nonlinear interaction is supposed to be
very weak, so that the field amplitude in a single pass through the crystal is only slightly affected. Therefore, the $z$ dependence of the
operators ${a}_{l}$ can be removed in Eq(\ref{Hamiltonian}). The longitudinal variation of the signal operator ${B}$ is due to the diffraction
described in the modal functions $f_{l}(\textbf{x},z)$.
\subsubsection {Near-field evolution}
At the mid-point plane $z=0$ of the crystal, designated as the near
field plane in the following, the ${B}$ field evolution can be
expressed as the sum of a damping and free propagation term inside the
cavity and of a parametric interaction term :
\begin{eqnarray}\label{evolution}
&&\frac{\partial {B}}{\partial
  t}(\textbf{x},0,t)=-\gamma(1+i\delta){B}(\textbf{x},0,t) \\
&&+ g \int\int d^2 x"
K_{int}(\textbf{x},\textbf{x"}){B}^{\dagger}(\textbf{x"},0,t)+\sqrt{2
  \gamma}{B}_{in}(\textbf{x},0,t)\nonumber
\end{eqnarray}
where $\gamma$ is the cavity escape rate, $\delta$ the normalized
cavity detuning of the modes, and ${B}_{in}$ the input field
operator. $K_{int}$ is the integral kernel describing the non-linear
interaction. Assuming exact collinear phase matching $k_p=2k_s$, and
neglecting walk off, this kernel associates two points $\textbf{x}$
and $\textbf{x"}$ through the pump amplitude at the average position
$\frac{\textbf{x}+\textbf{x"}}{2}$, and a function
$\Delta(\textbf{x}-\textbf{x"})$ describing the diffraction effects
within the crystal.

\begin{eqnarray}
K_{int}(\textbf{x},\textbf{x"}) &=&
A_p(\frac{\textbf{x}+\textbf{x"}}{2})
\Delta(\textbf{x}-\textbf{x"})\label{Kint}
\end{eqnarray}
with
\begin{equation}
\Delta( \textbf{x} - \textbf{x"} )=\frac{ik_s}{4 \pi l_c}
\int_{-l_c/2}^{l_c/2}\frac{dz'}{z'} e^{\frac{ik_s}{4z'}|\textbf{x} -
  \textbf{x"} |^2} \label{delta1}
\end{equation}

where $k_s= n_s \omega_s /c$ is the field wavenumber, and $n_s$ the
index of refraction at frequency $\omega_s$. It can be expressed in
terms of the integral sine function $Si(x)=\int_{0}^{x}\frac{\sin
udu}{u}$
\begin{equation}
\Delta( \textbf{x} - \textbf{x"} )=\frac{k_s}{2 \pi l_c}\left(
  \frac{\pi}{2} -Si(\frac{k_s|\textbf{x} - \textbf{x"} |^2}{2 l_c})
\right).
\end{equation}
In the thin crystal case ($ l_c\rightarrow 0$) the function
$\Delta( \textbf{x} \pm \textbf{x"} )$ tends to the usual
two-dimensional distribution $\delta(\textbf{x} \pm \textbf{x"}$.

In the thick crystal case, the parametric interaction mixes the
operators at different points of the transverse plane, over areas of
finite extension given by the spatial extension of the kernel
$K_{int}$. This extension is characterized by the width of the sine
function, which define a coherence length :
\begin{equation} \label{lcoherence} l_{coh}=\sqrt{\frac{\lambda l_c}
  {\pi n_s}}.
\end{equation}
When $|\textbf{x} - \textbf{x"} |\gg l_{coh}$, $\Delta$ and therefore
the kernel $K_{int}$ take negligible values, there is no coupling
between these two positions. On the other hand, when $|\textbf{x} -
\textbf{x"} |\ll l_{coh}$ the coupling mixes the fluctuations. Thus,
we can define $l_{coh}$ as the quantum resolution of our system.

Because of the finite size of the pump, the kernel will take
negligible values for $\textbf{x} + \textbf{x"}>w_p$.  Therefore, we
can define the number of transverse modes excited by the parametric
process inside the cavity, as the ratio between the size of the pump,
and the area defined by $l_{coh}$.

\begin{eqnarray}\label{nmode}
b=\frac{w_p^2}{l_{coh}^2}
\end{eqnarray}

This definition relies on the classical imaging properties of the
system. We will show in the last section of this article show that it is
consistent with the computation of the eigenmodes of the system.

\subsubsection {Far-field evolution}
\label{farmodel} Let us introduce the spatial Fourier transform of the
signal field envelope operator (\cite{Lopez})
\begin{eqnarray}
\tilde{B}(\textbf{q},z,t)&=&
\int\frac{d^{2}x}{2\pi}{B}(\textbf{x},z,t)e^{-i\textbf{q}\cdot\textbf{x}}
\end{eqnarray}
Equation (\ref{evolution}) becomes:
\begin{eqnarray}
&&\frac{\partial \tilde{B}}{\partial
  t}(\textbf{q},0,t)=-\gamma(1+i\Delta)\tilde{B}(\textbf{q},0,t)+ \\
&& g \int d^2 q"
\tilde{K}_{int}(\textbf{q},\textbf{q"}){B}^{\dagger}(\textbf{q"},0,t)+\sqrt{2
  \gamma}\tilde{B}_{in}(\textbf{q},0,t) \, , \nonumber
\label{evolution2}
\end{eqnarray}
where the coupling Kernel $\tilde{K}_{int}(\textbf{q},\textbf{q"})$ is
the Fourier transform of the kernel (\ref{Kint}) with respect to both
arguments.  Straightforward calculations show that
\begin{eqnarray}
\tilde{K}_{int}(\textbf{q},\textbf{q'})&=&
\tilde{A_{p}}(\textbf{q}+\textbf{q'}) {\rm
  sinc}[\frac{l_{c}}{2k_{s}}|\frac{\textbf{q}-\textbf{q'}}{2}|^2]
\label{Ktilde}
\end{eqnarray}
where $\tilde{A_p}$ is the spatial Fourier transform of the Gaussian
pump profile, i.e. $ \tilde{A}_{p}(\textbf{q}) = \frac{w_p^2}{2}A_p
\exp{(-|\textbf{q}|^2 \frac{ w_p^2}{4 })} $.

The sinc term in the coupling kernel of Eq. (\ref{Ktilde}) is the
Fourier transform of the $\Delta$ terms in Eq. (\ref{Kint})), and
correspond to the limited phase-matching bandwidth of the nonlinear
crystal. For a thin crystal, phase matching is irrelevant and there is
no limitation in the spatial bandwidth of down-converted modes,
whereas for a thick crystal, the cone of parametric fluorescence has
an aperture limited to a bandwidth of transverse wavevectors $\Delta q
\approx 1/l_{coh} \propto 1/ \sqrt{\lambda l_c}$. In the self imaging
geometry, the cavity ideally transmits all the Fourier modes, so that
the spatial bandwidth is only limited by the phase matching along the
crystal.  Finally, we have to notice that in the far field
configuration the $\tilde{A_{p}}(\textbf{q}+\textbf{q'})$ couples
different q-vectors modes within the finite width of the pump.


\subsection{Input/output relation}
In order to calculate the noise spectrum of the outgoing field, an
input/output method is used. The input field is supposed to be in a
coherent state and the fluctuations at the output can be inferred.
The relation linking the outgoing fields $B^{out}(\textbf{x},t)$ to
the intracavity and input fields at the cavity input/output
port\cite{Gardiner} is:
\begin{eqnarray}
B^{out}(\textbf{x},t)=\sqrt{2\gamma}B
(\textbf{x},t)-B^{in}(\textbf{x},t)\label{bound}
\end{eqnarray}
The evolution equation of the field, either in the near or in the far
field can be solved in the frequency domain by introducing:
$$
B^{in/out}(\textbf{x},\Omega)=\int\frac{dt}{\sqrt{2\pi}}B^{in/out}(\textbf{x},t)e^{-i\Omega
t }
$$
which lead to the input/output relation, linking
$B^{in}(\textbf{x},\Omega)$ and $B^{out}(\textbf{x},\Omega)$.

In the case of a thin crystal in the near field\cite{Petsas}, this
relation describes an infinite set of independent optical parametric
oscillators. In this case the squeezing spectrum can be calculated
analytically. More generally, this relation in near field links all
points in the transverse plane within the coherence area. In order to
get the input/output relation, we have to inverse the input/output
relation by using a numerical method used in \cite{Lopez}.

\subsection{Homodyne detection scheme in the near field and far field}
In the following sections we calculate the noise spectrum at the
output of the OPO as a function of the detected transverse mode
selected by an homodyne detection scheme \cite{homodynespatiale}. By
mixing it with a coherent Local Oscillator (LO) of various shape on a
50\% beamsplitter (reflection and transmission coefficients
$\textit{r}=\frac{1}{\sqrt{2}}$ and $\textit{t}=\frac{1}{\sqrt{2}}$),
one can measure the fluctuations on any transverse mode of the output
of the self imaging OPO, by measuring the photocurrents difference.
The two identical detectors of different size and position are
supposed to have a perfect quantum efficiency. All the fields are
evaluated at the beam-splitter location, and the z-dependence is
omitted in the following.

\begin{figure}[ht]
\centerline{ \scalebox{.5}{\includegraphics*{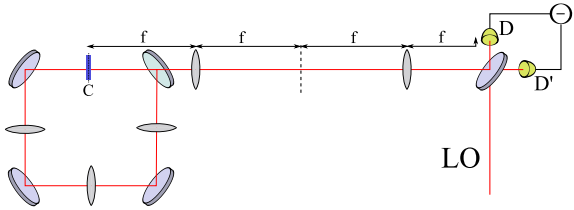}} }
\caption{Balanced homodyne detection scheme in the near field. Two
  matching lenses of focal f are used to image the cavity center C
  at the detection planes D and D'} \label{fig:fig4}
\end{figure}

\begin{figure}[ht]
\centerline{ \scalebox{.5}{\includegraphics*{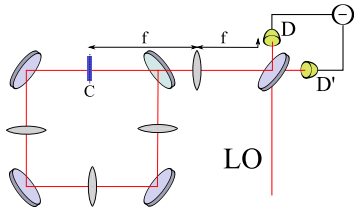}} }
\caption{Balanced homodyne detection scheme in the far field. A
  matching lens of focal f is used to obtain the far field image of
  the object plane C at the detection planes D and D'}
\label{fig:fig5}
\end{figure}

We use two different configurations : near-field (x-position basis) and far-field (q-vector basis). The complete detection scheme is
schematically shown in Fig. \ref{fig:fig4} and \ref{fig:fig5}. In the near-field configuration, the imaging scheme is composed of a two-lens
afocal system (focal length \textit{f}) which images the crystal/cavity center plane C onto the detection planes D and D'(near field planes). In
the far field configuration, a single lens of focal length \textit{f} transforms its focal object plane C into the image focal detection plane
D. Any image in the object plane C is tranformed into its fourier transform in the plane D (far field plane).

For near-field imaging, the local oscillator can be expressed as
$\alpha_{L}(\textbf{x},z)=|\alpha_{L}(\textbf{x},z)|e^{i\varphi_L(\textbf{x},z)}$.
The difference photocurrent is a measure of the quadrature operator:
\begin{eqnarray}\label{Nfhomodyne}
E_{H}(\Omega)=\!\!\!\!\int_{det}\!\!\!\! d\textbf{x}
\left[B^{out}(\textbf{x},\Omega)\alpha_{L}^{*}(\textbf{x})+
  B^{out+}(\textbf{x},-\Omega)\alpha_{L}(\textbf{x})\right]\,
\end{eqnarray}
where $\textit{det}$ is the image of the photodetection region at the crystal plane C, and assumed to be identical for the two photodetectors. %
The quantum efficiency of the photodetector is assumed to be equal 

For far-field imaging, the lens provides a spatial Fourier transform
of the output field $B_{out}(x,\Omega)$, so that at the location of
plane D the field $B^{out}_{D}(x,\Omega)$ is:
\begin{eqnarray}
B^{out}_{D}(x,\Omega)=\frac{2 \pi}{\lambda f}
\tilde{B}^{out}(\frac{2\pi}{\lambda f} x,\Omega)
\end{eqnarray}
In this plane, $B^{out}_{D}(x,\Omega)$ is mixed with an intense
stationary and coherent beam $\alpha_{LO}^{D}(x)=\frac{2\pi}{\lambda f
}\tilde{\alpha}_{LO}(\frac{2 \pi x }{\lambda f},\Omega)$, where
$\alpha_{L}(x)$ has a gaussian shape, with a waist $w_{LO}$. The
homodyne field has thus an expression similar to the near field case,
where functions of $x$ are now replaced by their spatial Fourier
transforms:
\begin{eqnarray}\label{FFHomodyne}
&&E_{H}(\Omega)= \\
&&\int_{det}d\textbf{q}
[\tilde{B}^{out}(\textbf{q},\Omega)\tilde{\alpha}_{LO}^{*}(\textbf{q})+
\tilde{B}^{out+}(\textbf{q},-\Omega)\tilde{\alpha}_{LO}(\textbf{q})]\nonumber
\end{eqnarray}

In near- and far-field, the fluctuations $\delta E_{H}(\Omega)$ of the
homodyne field around steady state are characterized by a noise
spectrum:
\begin{eqnarray}
V(\Omega)=\int_{-\infty}^{+\infty} d\Omega'\langle\delta
E_{H}(\Omega)\delta E_{H}(\Omega')\rangle=N+S(\Omega)
\end{eqnarray}
where $E_{H}$ is normalized so that $N$ gives the mean photon number
measured by the detector
\begin{eqnarray}
N=\int_{det} d\textbf{x} |\alpha_{L}(x)|^{2}
\end{eqnarray}
N represents the shot-noise level, and S is the normally ordered part
of the fluctuation spectrum, which accounts for the excess or decrease
of noise with respect to the standard quantum level (S=0). One should
note that there is a complete equivalence between a setup with a
finite and flat local oscillator and infinite detectors, and a flat
and infinite local oscillator combine with finite size detectors. We
will often use the configuration with finite size photodetectors in
the following.

\section{Non-classical properties}
\label{sec3}
\label{sec:effet}

We expose here the main properties of the fields emitted by the
sub-threshold self-imaging OPO. We will first consider the squeezing
in the near-field in a very similar manner as what was done for a
confocal OPO. Then we will study the far field properties and
demonstrate local EPR correlations.

\subsection{Squeezing in the near field}

As the self imaging cavity does not exert any spatial filtering on the
fields, the non-classical properties are very similar to those
observed in the single pass configuration.  We will here show the main
squeezing predictions for such a device, taking into account the
thickness of the crystal.  The corresponding calculations are
avalaible upon request to the authors.

\begin{figure}[ht]
\centerline{ \scalebox{.40}{\includegraphics*{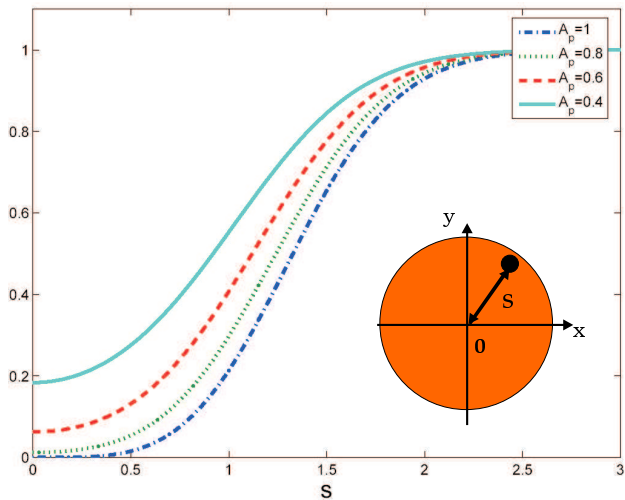}} }
\caption{ Quantum noise at zero-frequency normalized to the shot
  noise, for a pixel like detector located in the near field
  plane, as a function of the pixel distance from the origin $s$, normalized to the
  waist of the pump ($s=\frac{\rho}{w_p}$)  and for
  different pump values} \label{fig:fig6}
\end{figure}

Let us first consider the case of the thin crystal approximation, where no characteristic length is introduced in the model. We consider a thin
crystal self-imaging OPO pumped by a gaussian beam.  We look at the output quantum fluctuations with a pixel-like detector whose position is
varied. In figure \ref{fig:fig6}, the squeezing is plotted as a function of the detector distance from the optical axis for different mean
powers of the pump ($A_p=1$ corresponding to the threshold on the axis). The squeezing is maximum when the detection is centered on the pump
beam, and tends to zero far from the center. Figure \ref{fig:fig6} shows that the squeezing increases with the total pump power and depends
critically on its local value.  Hence, any transverse position on the crystal acts as an independent OPO.

\begin{figure}[ht]
\centerline{ \scalebox{.4}{\includegraphics*{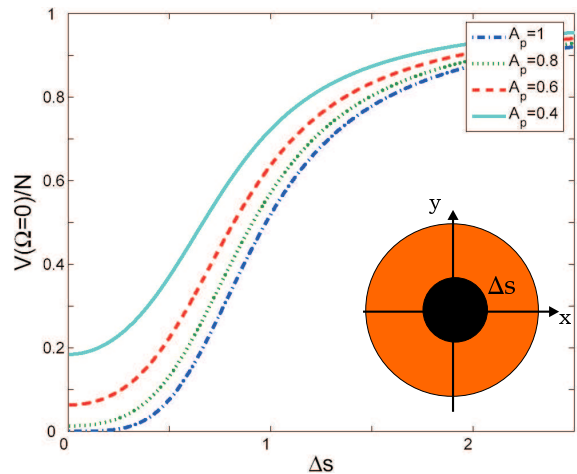}} }
\caption{ Quantum noise at zero-frequency, normalized to the shot
  noise, in the thin crystal case, for a circular detector in the near field plane, as a
  function of the radius of the detector normalized to the pump
  waist $\Delta s=\frac{\Delta \rho}{\omega_p}$ } \label{fig:fig7}
\end{figure}

The same behavior is observed in a configuration closer to actual
experimental scheme using a circular detector whose radius $\Delta
\rho$ can be varied. Like in the previous case, the curves in figure
\ref{fig:fig7} are crucially dependent on the pump power.

\begin{figure}[ht]
\centerline{ \scalebox{.4}{\includegraphics*{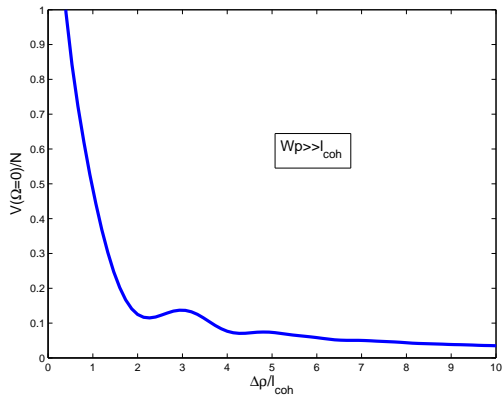}} }
\caption{ Quantum noise at zero-frequency, normalized to the shot
  noise, in the thick crystal case, as a function of the detector
  radius (scaled to $l_{coh}$).The detector in centered on the pump
  beam } \label{fig:fig8}
\end{figure}

\begin{figure}[ht]
\centerline{ \scalebox{.5}{\includegraphics*{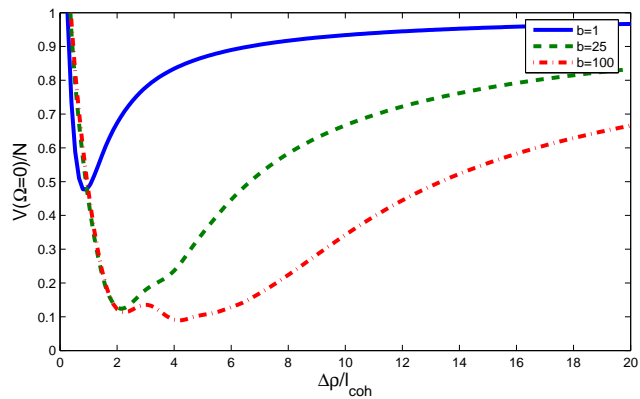}} }
\caption{Quantum noise at zero-frequency, normalised to the shot
  noise, in the thick crystal case, as a function of the radial
  size of the detector scaled to $ l_{coh} $, plotted for
  several values of the parameter
  $b=\frac{{w_p}^2}{{l_coh}^2}$} \label{fig:fig9}
\end{figure}

In a more general realistic study, we have to take into account the finite size of the crystal. For a thick crystal, the coherence length
$l_{coh}$ introduced in equation (\ref{lcoherence}) has to be taken into account. On transverse size smaller than this coherence length,
fluctuations are mixed inside the crystal.\\ In a first step, we consider a quasi plane pump for which its waist $w_p$, considered as infinite,
is much larger than the coherence area and can excite many modes. The detector is centered with respect to the optical axis and its size can be
changed. As shown in figure \ref{fig:fig8}, the squeezing is maximum when its size is larger than the coherence area, and the noise tends to
shot noise for a pixel like detector, whose size becomes smaller than the coherence length. At the scale of the coherence area, the OPO can be
considered as locally single mode and the local fluctuations are mixed. In regions smaller than the coherence area independent modes having
their own fluctuations cannot be excited. For large detectors, several coherence area can be excited, the OPO can be considered as multimode,
and the squeezing is maximum.  The coherence length sets the limit between single mode and multimode operation.\\ In a second step, we have to
consider a more realistic case for which the pump waist $w_p$ is finite.  In figure \ref{fig:fig9}, we represent the quantum noise as a function
of the detector size for different pump waist (normalized to the coherence length). For detectors smaller than the coherence area, the quantum
noise goes to shot noise whatever the size of the pump is. As explained in the last paragraph, at that scale the OPO can be considered as
locally single mode and no squeezing can be obtained. For detectors whose size is close the coherence area one, squeezing is obtained.
Nevertheless this squeezing degrades for a given pump waist when the size of the detector increases. For a given pump size, when the detector
becomes larger than the excited surface, vacuum fluctuations are coupled to the detected signal and the squeezing degrades. In the same way, for
a given detector size, the squeezing decreases with the waist of the pump. In fact a finite pump size limits the number of excited modes.
Increasing the pump size, increase the number of excited modes and improve the squeezing.

\subsection{Entanglement in the far field}


In the far field, the analysis has to be performed not in crystal
plane (near field plane) but in its Fourier plane (far field
plane). Squeezing can be observed in the far field when using a
symmetric detector. Indeed, contrary to the near-field case, in the
far field configuration the down conversion process couples two
symmetric $k$ vectors. Thus in order to recover the squeezing one
needs a symmetric detector relative to the optical axis of the imaging
system. The results obtained are therefore the
same as those in the confocal cavity, both with a plane pump and with
a finite pump. Corresponding calculations are also available upon
request to the authors.

The advantage of the self imaging cavity is that it does not couple
the two symmetrical $k$ vectors. Therefore one expects correlations
between two symmetrical areas in the far field, as it will be shown in
the following.

\begin{figure}[ht]
\centerline{ \scalebox{.9}{\includegraphics*{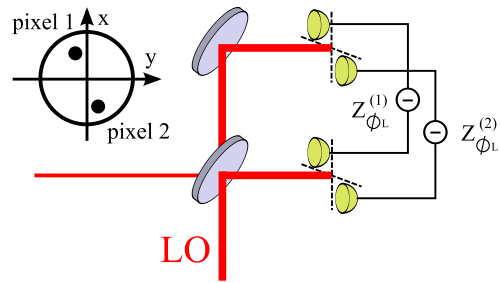}} }
\caption{Homodyne detection scheme for the measurement of the
  quadrature components of the output field on two symmetrical pixels:
  pixel 1 and pixel 2} \label{fig:fig11}
\end{figure}

In order to characterize the correlation level between symmetrical
parts of the beam, we compare the quadrature field fluctuations on two
symmetrical pixels. In order to get this quantities, we use the
homodyne detection scheme (figure \ref{fig:fig11}) proposed in
\cite{homodynespatiale}, where two symmetrical sets of two detectors measure a
quadrature of the field at two symmetrical positions.

Let us consider a pixel-like detector with finite detection area
$\Delta\rho_{j}$, according to equation \ref{FFHomodyne} the detected
field quadrature is given by :
\begin{eqnarray}
E_{\phi_{L}}^{(j)}(\Omega) & = & \int_{\Delta\rho_{j}}\!\!\!\!\!\!
d\textbf{q}
\left[\tilde{B}^{out}(\textbf{q},\Omega)|\alpha_{L}(\textbf{q})|e^{-i\phi_L}+\right.
  \nonumber \\
& & \left.\tilde{B}^{out+}(\textbf{q},-\Omega)|\alpha_{L}(\textbf{q})|e^{i\phi_L}\right] 
\end{eqnarray}
where we have introduced explicitly the phase of the local oscillator.
To compare the fluctuations of the field quadrature measured in two
symmetrical pixels $j=1$ and $j=2$, we compare the sum and the
difference of these quantities.
\begin{eqnarray}
E_{\phi_{L}}^{(\pm)}(\Omega)=E_{\phi_{L}}^{(1)}(\Omega)\pm
E_{\phi_{L}}^{(2)}(\Omega)
\end{eqnarray}
In order to evaluate the degree of correlation or anti-correlation, we
introduce the corresponding fluctuations spectra:
\begin{eqnarray}\label{correlation}
V_{\phi_{L}}^{(\pm)}(\Omega)=\int_{-\infty}^{+\infty}d\Omega'\langle
E_{\phi_{L}}^{(\pm)}(\Omega) E_{\phi_{L}}^{(\pm)}(\Omega') \rangle
\end{eqnarray}
Straightforward calculations show that:
\begin{eqnarray}
V_{\phi_{L}}^{(-)}(\Omega)=V_{\phi_{L}+\pi/2}^{(+)}(\Omega)
\end{eqnarray}
It results that the correlation between $E_{\phi_{L}}^{(1)}$ and
$E_{\phi_{L}}^{(2)}$ is the same that the anticorrelation between the
corresponding orthogonal quadrature components
$E_{\phi_{L}+\pi/2}^{(1)}$and$E_{\phi_{L}+\pi/2}^{(2)}$. In order to
calculate (\ref{correlation}), we develop the expression, so as:
\begin{eqnarray}
V_{\phi_{L}}^{(\pm)}(\Omega)&=&\int_{-\infty}^{+\infty}d\Omega'\langle
E_{\phi_{L}}^{(1)}(\Omega) E_{\phi_{L}}^{(1)}(\Omega') \rangle\nonumber \\
&&+\int_{-\infty}^{+\infty}d\Omega'\langle
E_{\phi_{L}}^{(2)}(\Omega) E_{\phi_{L}}^{(2)}(\Omega') \rangle\nonumber \\
&&\pm\int_{-\infty}^{+\infty}d\Omega'\langle
E_{\phi_{L}}^{(1)}(\Omega) E_{\phi_{L}}^{(2)}(\Omega') \rangle\nonumber \\
&&\pm\int_{-\infty}^{+\infty}d\Omega'\langle
E_{\phi_{L}}^{(2)}(\Omega) E_{\phi_{L}}^{(1)}(\Omega') \rangle
\end{eqnarray}
The terms $\int_{-\infty}^{+\infty}d\Omega'\langle
E_{\phi_{L}}^{(i)}(\Omega) E_{\phi_{L}}^{(i)}(\Omega') \rangle$
correspond to the result of the fluctuation spectra of a homodyne
detection scheme using a single pixel. The other terms are cross
correlation terms, so that:
\begin{eqnarray}\label{EPR}
V_{\phi_{L}}^{-}(\Omega)=V_{\phi_{L}+\pi/2}^{+}(\Omega)
\end{eqnarray}

When these variances are bellow one, EPR beams are obtained at
the output of the self-imaging cavity. One should note that these
variances correspond to the fluctuation spectrum obtained performing a
homodyne detection in the far field with symmetric detectors : the usual
connection between squeezing and quantum correlations is exhibited, both
side of the same phenomenon\cite{Quantim}. More specifically, the spatial entanglement in
the far field arise from the correlations between the modes
${a}_{\textbf{q}} \sim e^{i\textbf{q.x}}$ and ${a}_{-\textbf{q}} \sim
e^{-i\textbf{q.x}}$. As ${a}_{\textbf{q}}$ and ${a}_{-\textbf{q}}$ are
EPR entangled beams, it is well know that the combination of modes:
\begin{eqnarray}
\frac{{a}_{\textbf{q}}+{a}_{-\textbf{q}}}{\sqrt{2}} \sim
\cos(\textbf{q.x})
\nonumber\\
\frac{{a}_{\textbf{q}}-{a}_{-\textbf{q}}}{\sqrt{2}} \sim
\sin(\textbf{q.x})
\end{eqnarray}
will be squeezed with respect to two orthogonal quadrature components.
The modes proportional to $\cos(\textbf{q.x})$ are the even modes:
using an even detection scheme it is possible to see squeezing, as
already found in the previous section. Note that if we use an odd
detection scheme (symmetrical detectors with an odd local oscillator),
it will be also possible to see squeezing in the far field, but on the
orthogonal quadrature.

In order to ascertain the inseparable character of this physical state,
Duan et al.\cite{Duan} have shown one needs to make two joint
correlation measurements on non commuting observables on the
system. They have shown that in the case of gaussian states there
exists a criterion of separability in terms
of the quantity $S_{12}$, that we will call 'separability', and is
given by:
\begin{eqnarray}\label{cduan}
S_{12}(\Omega)=\frac{1}{2}(V_{\phi_{L}}^{-}(\Omega)+
V_{\phi_{L}+\frac{\pi}{2}}^{+}(\Omega))
\end{eqnarray}
The suddicient Duan criterion for inseparability is given by:
\begin{eqnarray}
S_{12}(\Omega)<1
\end{eqnarray}

\begin{figure}[ht]
\centerline{ \scalebox{.5}{\includegraphics*{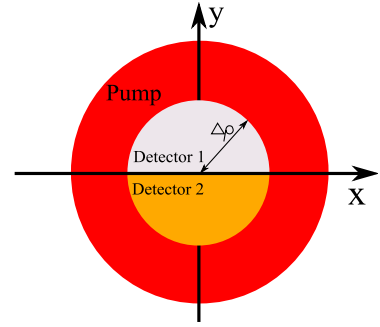}} }
\caption{Detection scheme for the inseparability measurement }
\label{fig:fig12}
\end{figure}

\begin{figure}[ht]
\centerline{ \scalebox{.4}{\includegraphics*{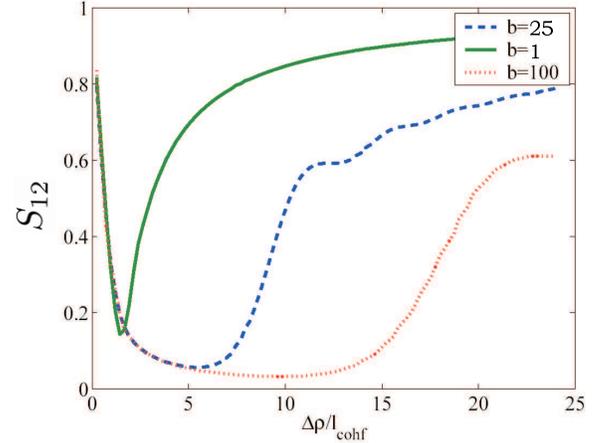}} }
\caption{Inseparability at zero-frequency, and at resonance, as a
  function of the radial amplitude of the detector
  $\Delta\rho$(scaled to the coherence area $l_{cohf}$)
  , in the finite pump regime
  and far field approach and for different values of
  b.} \label{fig:fig13}
\end{figure}

First, we can perform a joint correlation measurement using two
split detectors of same but variable size as depicted in
Figure \ref{fig:fig12}. Figure \ref{fig:fig13} shows the evolution of the separability at zero
frequency $S_{12}(0)=S_{12}$ for different b parameters, in function
of the detector radius scaled to $l_{cohf}=\lambda f/2\pi w_p$ \cite{Lopez}. Notice
that results are the same as a local squeezing measurement using
a circular detector of variable radius $\Delta\rho$ centered on the
optical axis.

\begin{figure}[ht]
\centerline{ \scalebox{.4}{\includegraphics*{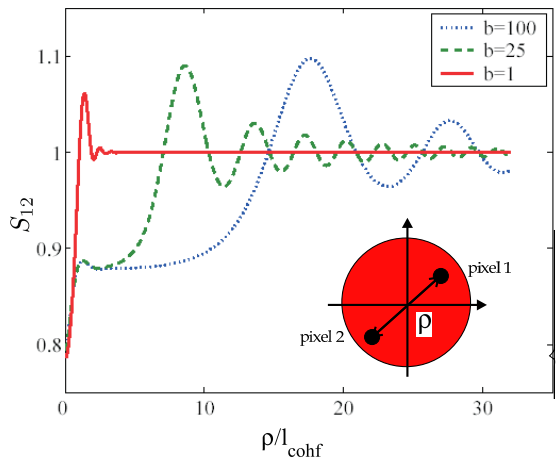}} }
\caption{Inseparability at zero-frequency, and at resonance, as a
  function of the distance between the two pixels $\rho$ (scaled to
  the coherence area $l_{cohf}$, in the
  finite pump regime and far field approach and for different values
  of b.}
\label{fig:fig14}
\end{figure}

Fig. \ref{fig:fig14} shows the results obtained in the case of two
symmetrical pixels (pixel of size equal to the coherence length
$l_{cohf}$, for different b values, in
function of the distance between the two pixels $\rho$.


\section{Pixel-based model for the self-imaging OPO}
\label{sec4}

We have so far described the non-classical properties of the
self-imaging OPO using detection based geometry, very appropriate to
describe actual experiments. However, it is known that any
input/output system can be described by eigenmodes : for instance in
the case of mode-locked pulses of light incident on a
non-linear cristal, it has been shown that independent modes could be
found, either using the Schmidt decomposition in the single photon
regime \cite{Walmsley} or diagonalising the coupling matrix in the
continuous wave regime \cite{Patera}. We propose here to use the same
technique to exhibit the eigenmodes of the system and give a more
precise value of the number of modes involved in the process.


Let us pixelize the transverse space with pixels much smaller than
$l_{coh}$ and develop the OPO equations onto the pixel operators. To
simplify the system, we can first consider a one dimension
pixelization. Let $L$ be the size of the pixelized zone, and $N$ the
number of pixels. The pixel $i$ is defined as the zone of size
$\frac{L}{N}$ near the abscissa $x_i=i\frac{L}{N}$, with i ranging
from $-N/2$ to $N/2$. The pixel operator is therefore :

\begin{equation}
B_i = \int_{S_i} dx B(x)
\end{equation}

The pixel size must be chosen small enough to ensure the constant
value of $A_p$ and $K_{int}$ on every pixel. In this case, the Kernel
can be written as :

\begin{equation}
K_{int}(i,j)=K_{int}(\textbf{x}_i,\textbf{x}_j)
\end{equation}

and the evolution equation (\ref{evolution}) at zero frequency then
becomes :

\begin{equation}\label{equationpixel}
-\gamma{B}_i + \frac{gL}{N} \sum_j K_{int}(i,j){B}^{\dagger}_j+\sqrt{2\gamma}{B}^{in}_i=0
\end{equation}

\begin{figure}
\begin{center}
  \subfigure[ ]{\includegraphics[width=6cm]{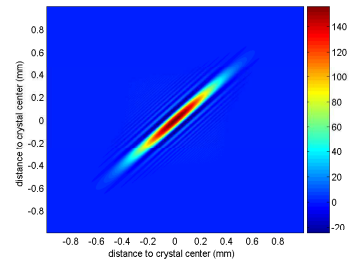}}
  \\
  \subfigure[ ]{\includegraphics[width=9cm]{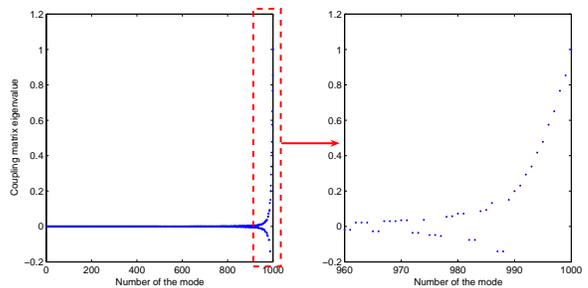}} \caption{a)
    Coupling Matrix $K_{int}$ between two points of the crystal b) Spectrum of this
    matrix. }
\label{KMat}
\end{center}
\end{figure}

\begin{figure}
\includegraphics[width=9cm]{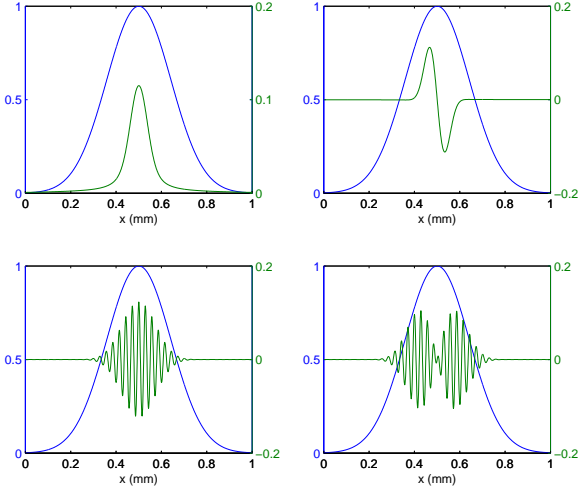}
\caption{Shape of the eigenvectors of matrix $K_{int}$ (green) for
  the two highest positive (top) and negative (bottom) eigenvalues
  compared to the one of the pump (blue)}
\label{modespropres}
\end{figure}

To solve these N coupled equations, one must find the eigenvectors and
eigenvalues of the matrix
$K_{int}(i,j)$. 
The K matrix and its spectrum are represented on figure
\ref{KMat}. Its diagonalization gives a set of eigenmodes with
corresponding eigenvalues. Some of these eigenmodes are represented
on figure \ref{modespropres}, they are very close to Hermite-Gauss
polynomials shapes whose characteristic waist is imposed, in our case,
by the pump waist.

These modes form a basis of uncorrelated modes of the emitted
light. Indeed, let us call $C_k$ the eigenmode of eigenvalue $\lambda_k$. As
$K_{int}(i,j)$ is both self-adjoint and real, $\lambda_k$ and $C_k$
components are all real. In this basis, equation \ref{equationpixel}
can be rewritten as set of equations, one per mode :

\begin{equation}\label{equationpropres}
-\gamma{C}_k + \frac{gL}{N} \lambda_k {C}^{\dagger}_k+\sqrt{2\gamma}{C}^{in}_k=0
\end{equation}

These equations can again be decoupled, using the quadrature operators
:

\begin{eqnarray}
C_{k+}= C_k+C_k^\dagger\, \, \,\\
C_{k-}= -i (C_k-C_k^\dagger)
\end{eqnarray}

The final set of equations is now given by

\begin{eqnarray}
-\gamma C_{k+} + \frac{gL}{N} \lambda_k C_{k+} + \sqrt{2\gamma} C_{k+}^{in} = 0 \\
-\gamma C_{k-} - \frac{gL}{N} \lambda_k C_{k-} + \sqrt{2\gamma} C_{k-}^{in} = 0
\end{eqnarray}

In this basis, using the input/output relations \ref{bound}, we can
calculate the squeezing properties of the modes $C_{k\pm}^{out}$, in
the near field of the $z=0$ plane. Using the same method as in
\cite{Patera}, the fluctuations at zero frequency of the quadratures
of the eigenmode $C_{k\pm}^{out}$, normalized to the shot-noise level,
are given by :

\begin{eqnarray}
V_{k\pm}=\Lambda_k^\pm = \frac{1\mp r\frac{\lambda_k}{\lambda_{max}}}{1\pm r\frac{\lambda_k}{\lambda_{max}}}
\end{eqnarray}

where $r$ is the pump power normalized to the threshold and
$\lambda_{max}=\max_k \Lambda_k$ the highest eigenvalue of
$K_{int}$. $\lambda_{max}$ is of special interest since it is
related to the pump power at threshold and $C_{max}$ is the corresponding
lasing mode. One can see in the previous equation that for each mode
whose eigenvalue is different from zero one of its two variances is
bellow one, implying that it is non-classical. However,
for eigenvalues very small compared to $\lambda_{max}$, the squeezing
is negligible. Thus one can compute the number of relevant mode of
the system, for instance using a threshold eigenvalue (about 10\% of
the maximum eigenvalue). Another possibility is to calculate the
cooperativity \cite{Walmsley}, defined from the eigenvalues of the matrix.

\begin{eqnarray}
\kappa=\frac{(\sum \lambda _k ^2)^2}{\sum \lambda_k ^4}
\end{eqnarray}

The obtained number of modes is very close to the one defined in
equation (\ref{nmode}) in a 1D case. For example, using
typical experimental values (1 cm long crystal of index 2 and a 300
$\mu m$ at 1064 nm), we find $b=\frac{w_p}{l_{coh}}= 7.5$ and
$\kappa=6.8$. This means that in the 2D case, our
self-imaging OPO can potentially excite 50 modes.

One should note that from these eigenmodes it is possible to find the
noise properties of the pixel operators $B_i^{out}$ after the cavity, in
the near field of the crystal, by inverting the $K_{int}$ matrix,
which gives :

\begin{eqnarray}
B_{i\pm}^{out}&=&\sum_k V_{ik} C_{k\pm}^{out}=\sum_k V_{ik} \Lambda_k^\pm C_{k\pm}^{in} \\
&=&\sum_{kj} V_{ik}V_{jk} \Lambda_k^\pm B_{j\pm}^{in}
\end{eqnarray}

Using these expressions, we can calculate the measured fluctuations of
the quadratures on a detector with an arbitrary shape :

\begin{eqnarray}
V_{det\pm}=\frac{\sum_{ij\in det} <B_{i\pm}^{out} B_{j\pm}^{out}>}{\sum_{i\in det} <(B_{i\pm}^{out})^2>}
\end{eqnarray}
These numerical simulations show the exact same results as the
analytical results presented in section \ref{sec3}.

We thus have shown two ways of solving the problem, each having
different physical significance. Indeed, in the approach of section
\ref{sec3} we have seen that the system has a coherence area that
defines the smallest mode having non-classical properties. This is
relevant of quantum imaging applications as it gives which pixel size
one can address with quantum techniques. In the present section we
have shown that a proper description of the system consists of an
eigenmodes decomposition, modes that have Hermite-Gauss shape and
whose squeezing decreases with the mode number. However, these modes
shape are complex to measure experimentally.\\




Laboratoire Kastler-Brossel, of the Ecole Normale Sup\'erieure and the
Universit\'{e} Pierre et Marie Curie - Paris 6, is associated with the Centre
National de la Recherche Scientifique. We acknowledge the financial
support of the Future and Emerging Technologies (FET) programme within
the Seventh Framework Programme for Research of the European
Commission, under the FET-Open grant agreement HIDEAS, number
FP7-ICT-221906
\\

\end{document}